\documentclass[12pt,a4paper]{article}

\usepackage{enumitem}
\usepackage{CJKutf8}
\usepackage[T1]{fontenc} 
\usepackage{tgtermes} 

\usepackage{tabularray} 
\UseTblrLibrary{booktabs} 

\usepackage{graphicx} 
\usepackage[margin=25mm]{geometry}
\usepackage{amsmath}
\usepackage{amsfonts}
\usepackage{amssymb}
\usepackage{siunitx}
\usepackage{verbatim}
\usepackage{natbib} 
\usepackage{microtype} 

\usepackage{natbib} 
\usepackage{microtype} 

\usepackage[hyphens]{url}
\usepackage[hidelinks]{hyperref}
\usepackage[hyphenbreaks]{breakurl}

\usepackage[hyphens]{url}
\hypersetup{colorlinks=true, linkcolor=blue, urlcolor=blue, breaklinks=true}

\author{}
\date{}

\title{US-China perspectives on extreme AI risks and global governance
}
\author{
  Akash R. Wasil\footnote{Author correspondence: aw1404@georgetown.edu}\\
  \small{Georgetown University}\\
  \and 
  Tim Durgin\\
   \small{Independent}\\
}\date{}

\begin{document}
\begin{CJK}{UTF8}{gbsn}
\maketitle

\begin{abstract}

\noindent
The United States and China will play an important role in navigating safety and security challenges relating to advanced artificial intelligence. We sought to better understand how experts in each country describe safety and security threats from advanced artificial intelligence, extreme risks from AI, and the potential for international cooperation. Specifically, we compiled publicly-available statements from major technical and policy leaders in both the United States and China. We focused our analysis on advanced forms of artificial intelligence, such as artificial general intelligence (AGI), that may have the most significant impacts on national and global security. Experts in both countries expressed concern about risks from AGI, risks from intelligence explosions, and risks from AI systems that escape human control. Both countries have also launched initial efforts designed to promote international cooperation around safety standards and risk management practices. Notably, our findings only reflect information from publicly available sources. Nonetheless, our findings can inform policymakers and researchers about the state of AI discourse in the US and China. We hope such work can contribute to policy discussions around advanced AI, its global security threats, and potential international dialogues or agreements to mitigate such threats.

\end{abstract}
\newpage

\section{Introduction}\label{introduction}

\textbf{Artificial intelligence is a transformative technology with major implications for national and global security.} Many AI experts have expressed concerns about AI-related national and global security threats. Examples include risks from AI-enabled biological weapons, risks from autonomous AI systems that escape human control, and risks from AI applied in military operations (Bengio et al., 2024; Hendrycks et al., 2023).

\textbf{The United States and China are the world's leaders in AI development.} In 2022, the size of the AI market in the US amounted to USD 103.7B (Statista-1, 2024), while the size of the Chinese AI market was approximately USD 40B (Statista-2, 2024). As of 2022, 26 percent of the world's top AI researchers came from China, while 28 percent came from the United States (Yang 2024). US companies dominate the top of the large language model (LLM) rankings: examples include Open AI's Chat-GPT, Anthropic's Claude, and Google's Gemini (Guinness, 2024). While most Chinese LLMs lag quite a bit behind, there has been progress in recent years. Specifically, Moonshot AI's Kimi model, under certain conditions, can achieve performance comparable to Chat GPT-4 (Zhang, 2023).

\textbf{We aimed to acquire a better understanding of AI policy discourse in China and the United States.} Discourse about US-China relations is often partisan, rather than grounded in objective data. Discussions about US-China AI policy and potential international agreements should be rooted in a concrete understanding of attitudes toward AI, safety and security, and international cooperation in both countries. To improve our understanding of these topics, we compiled publicly available statements from Chinese and US officials regarding artificial intelligence, safety and security risks, and international cooperation and global governance. We focus our analysis on statements regarding the most powerful forms of artificial intelligence (e.g., artificial general intelligence or“AGI") and the most extreme risks posed by AI (e.g., developing biological weapons, developing novel kinds of weapons of mass destruction, or escaping human control.) 

\section{Methodology}\label{Methodology}

We included publicly available material about extreme AI risks, advanced AI, safety and security challenges, and international governance. Our approach was semi-structured and not meant to be systematic or comprehensive. For both searches, we focused on statements that acknowledged extreme risks from AI or AI safety and security threats– notably, our aim was not to characterize debates about the likelihood or plausibility of such risks. Below, we offer more details about our approach for both the China search and the US search.

When searching for statements by Chinese leaders, we first identified commonly used terms for extreme AI risks. We started with 通用人工智能, the most commonly used word in Chinese for General Artificial Intelligence. Using searches with this keyword, we identified additional keywords such as 生存风险 (existential risk) and 对齐 (alignment). We prioritized the inclusion of statements from government officials and officials from high-ranking academic institutions that regularly advise government officials. When we identified a relevant statement with an English translation, we used the English translation. Otherwise, the 2nd author (who is fluent in Chinese) translated the statements. Since our analysis was focused on Mainland China, we searched for the keyword terms in Simplified Chinese. When searching for statements by US leaders, we followed a similar procedure. We focused on statements from policymakers or statements given to government bodies (e.g., testimony to Congressional hearings). 

When determining which statements to include, we prioritized statements by senior faculty at top universities, experts working at frontier AI companies, policymakers and government officials, and officials who would likely advise policymakers. We were also most interested in characterizing statements relating to extreme risks related to advanced AI, safety and security, and the potential for international cooperation or global governance.

\section{Results}\label{Results}

\subsection{China}
\subsubsection{Safety and security}
When examining the discourse among Chinese scholars and policymakers, there is a clear acknowledgment of long-term security risks from advanced AI development. The discourse includes descriptions of specific safety and security concerns– such as intelligence explosions, the potential for self-replication and deception, and the importance of AGI security. Chinese government officials are concerned about AI's potential to generate “unhealthy" or “illegal" information. Finally, some influential minority of Chinese scholars have expressed concerns about potential existential risks created by advanced AI.

Example quotes (bolding added):

\begin{itemize}
\item 
     “If there is an \textbf{intelligence explosion} once AI has evolved to a certain level, the default result will inevitably be catastrophic.” – Paper by Scientists from Peking University's Dept. of Computer Science\footnote{Peking University is one of the top two universities in China.}

\item 
    “[Referring to AI] Every technology is like nuclear technology– if humankind had a choice, \textbf{perhaps it would be best not to discover dangerous materials like uranium and radium}.” – Ya-Qin Zhang, director of the Tsinghua Institute for AI Industry Research (Zhang, Ya-Qin, 2023)

\item 
    “Service providers must pay close attention to the generative AI's potential long-term risks. They must treat with caution any artificial intelligence (models) with the \textbf{ability to deceive humans, self-replicate, or self-modify}.” – Gen. AI Security Requirements instituted by the National Information Security Standardization Technical Committee of China (TC260)\footnote{The National Information Security Standardization Technical Committee of China (often abbreviated as TC260) is China's leading organization for writing national standards related to cybersecurity.} (TC260, 2024)

\item 
    “(AI systems) must use keywords, classification models, and other methods to monitor user input. If the user makes 3 inputs in a row that contain \textbf{illegal, unhealthy informatio} and/or are obviously designed to induce the AI system to generate unhealthy information, or if the user makes a total of 5 such inputs in one day, approved, legal methods shall be used to suspend said user's (AI) services.”– TC260 Gen. AI Security Requirements (TC260, 2024)

\item 
    “Main security Risks of AI Corpus and AI Generated Content: Content that \textbf{violates core Socialist values}, including content which: a) incites the subversion of state power and/or the overthrow of the Socialist System; b) harms national security, the national interest, and/or the nation's image; c) incites the dissolution of the state, damages national unity and/or social stability; d) promotes terrorism and/or other extremist ideologies.” – TC260 Gen. AI Security Requirements (TC260, 2024)

\item 
     “If AIA (AI Actants) evolve to the point where they need to defend their \textbf{“right to survive"} and do not hesitate to harm human beings in order to protect themselves, they will become autonomous systems that can fight against human beings, which will lead to disasters for humanity.” – Bingxing Fang, Director of Guangzhou University's Network Security Technology National Engineering Laboratory (Fang, 2020)

\item 
    "The superintelligence of the future may see humans as humans see ants today, and if humans can't treat other types of life with kindness, \textbf{why should the superintelligence of the future treat humans with kindness?} The biggest bottleneck to whether humans and AI can coexist in the future lies in humans, not AI.” – Yi Zeng, Director of the International Research Center for AI Ethics and Governance at the Chinese Academy of Sciences (Zeng 2023).

\end{itemize}

\subsubsection{International cooperation and global governance}

When examining discourse on international cooperation and global governance in China, we observed several expressions of interest in global cooperation on AI and risk management. More specifically, the Chinese government has expressed a desire to promote tiered testing systems with different requirements based on AI risk levels. Additionally, China is very concerned about ensuring equal access to AI technology across global borders.

Example quotes:
\begin{itemize}
\item 
    “We should promote the establishment of a \textbf{testing and assessment system based on AI risk levels}, implement agile governance, and carry out tiered and category-based management for rapid and effective response.” – China's AI Global Governance Initiative (Embassy of the People's Republic of China, 2023)

\item 
    “We support discussions within the United Nations framework to establish an \textbf{international institution to govern AI}.” – China's AI Global Governance Initiative (Embassy of the People's Republic of China, 2023)

\item 
    “We call for \textbf{global collaboration to foster the sound development of AI}, share AI knowledge, and make AI technologies available to the public under open-source terms.” – China's AI Global Governance Initiative (Embassy of the People's Republic of China, 2023)
    
\item 
    “AI must be fair. All countries should be able to participate in AI governance and \textbf{share in its benefits; no country should be left behind}." – China's AI Global Governance Initiative (Embassy of the People's Republic of China, 2023)

\item 
    “AI research and governance need the \textbf{participation and cooperation of people in different fields all over the world}. In addition to those engaged in AI R\&D and use, they also need the participation of people in different fields such as law, morality and ethics. We need to clarify the standards of ethics and morality and what it means to be ‘moral' and ‘ethical'.” – Bo Zhang Professor, Tsinghua University's Department of Computer Science\footnote{Tsinghua University is one of the top 2 universities in China.} (Zhang, 2022)
    
\end{itemize}

\subsection{United States}
\subsubsection{Safety and Security}
Discourse among American scholars and policymakers also includes clear acknowledgment of long-term security risks from advanced AI development. AI experts in the US have expressed strong concern about potential catastrophic outcomes from advanced AI. More recently, policymakers have begun to discuss artificial general intelligence and its risks. Many US policymakers and AI experts recognize the potential for extreme risks, both from malicious use of advanced AI and from potential loss of control scenarios. US policymakers have also expressed an interest in learning more about artificial general intelligence, standardizing the definition of AGI, understanding the likelihood and magnitude of AGI risks, and implementing policies to prepare for AGI risks. 

Example quotes:
\begin{itemize}
\item 
    “There is a spectrum of problems we could face related to this, at the extreme end of which [are] concerns about whether a sufficiently powerful AI, without appropriate safeguards, could be a threat to humanity as a whole–referred to as existential risk. \textbf{Left unchecked, highly autonomous, intelligent systems could also be misused or simply make catastrophic mistakes}.” – Dario Amodei, testifying before the Senate Judiciary Committee (Amodei, 2023).
\item 
    “The third possibility, which could emerge in as little as a few years, is that of \textbf{loss of control}, when an AI is given a goal that includes or implies maintenance of its own agency, which is equivalent to a survival objective... If the AI is misspecified, powerful enough, and exploits a loophole in its goals, the consequences could be unforeseen and severe. Therefore, a reactive approach to mitigating misspecified goals could be extremely costly for society, and we may only have a few chances of getting the alignment right for superhuman AI.” – Yoshua Bengio, testifying before the Senate Judiciary Committee (Bengio, 2023).
\item 
    “As long as there are really thoughtful people, like Dr. Hinton or others, who worry about the \textbf{existential risks of artificial intelligence—the end of humanity—I don't think we can afford to ignore that}... Even if there's just a one in a 1000 chance, one in a 1000 happens. We see it with hurricanes and storms all the time.” – Representative Beyer (Henshall, 2024). 

\item 
    “The AI Working Group recognizes that there is not widespread agreement on the definition of AGI or threshold by which it will officially be achieved. Therefore, we encourage the relevant committees to better define AGI in consultation with experts, \textbf{characterize both the likelihood of AGI development and the magnitude of the risks that AGI development would pose}, and develop an appropriate policy framework based on that analysis.” – Senators Schumer, Rounds, Heinrich, and Young (Bipartisan Senate AI Working Group, 2024).
\item 
    “Artificial intelligence (AI) has the potential to dramatically improve and transform our way of life, but also presents a spectrum of risks that could be harmful to the American public, some of which could have catastrophic effects. Extremely powerful frontier AI could be misused by foreign adversaries, terrorists, and less sophisticated bad actors to cause widespread harm and threaten U.S. national security. Experts from across the U.S. government, industry, and academia believe that advanced AI could one day enable or assist in the \textbf{development of biological, chemical, cyber, or nuclear weapons}.” – Senators Romney, Reed, Moran, and King in the Framework for Mitigating Extreme AI Risks (Romney et al., 2024). 
\end{itemize}

\subsubsection{International cooperation and global governance}

When examining discourse in the United States around international cooperation and global governance, we observed an interest in international cooperation on AI standards and risk management practices. For example, a recent bill has outlined specific ways that the US can cooperate with international partners on safety standards, and the recently established US AI Safety Institute has highlighted international cooperation in its strategic vision. 

Example quotes:
\begin{itemize}
\item 
    “[This bill] directs the heads of Commerce, State and the Office of Science and Technology Policy (“OSTP”) to jointly seek to \textbf{form alliances or coalition with like-minded countries to cooperate on approaches to innovation in AI and to coordinate and promote the development and adoption of common AI standards}... [the bill] ensures that participating countries maintain adequate research security measures, intellectual property protections, safety standards, and risk management approaches.” – Bill summary of the Future of AI Innovation Act (Senate Committee on Commerce, Science, and Transportation, 2024).
\item 
    “[The US AI Safety Institute aims to] lead an \textbf{inclusive, international network on the science of AI safety}. AI safety practices must be globally adopted to the greatest extent possible. We intend to serve as a partner for other AI Safety Institutes, national research organizations, and multilateral entities like the OECD and G7. We intend to work with our partners to foster commonly accepted scientific methodologies with the intention of developing a shared and interoperable suite of AI safety evaluations and agreed-upon risk mitigations. In doing so, we hope to help develop the science and practices that underpin future arrangements for international AI governance.” – US AI Safety Institute (NIST, 2024)
\end{itemize}

\subsection{Joint statements}
\subsubsection{Safety and security}

Joint statements including representatives from the United States and China have acknowledged safety and security threats from AI. Several of these statements were made in the context of international summits or track II dialogues. 

Example quotes:
\begin{itemize}
\item 
    “\textbf{Particular safety risks arise at the frontier of AI}. Substantial risks may arise from potential intentional misuse or unintended issues of control relating to alignment with human intent.” (Bletchley Declaration, 2023)\footnote{The Bletchley Declaration was signed jointly by the governments of the United States, China, and 27 other nations.}
    
\item 
    “\textbf{Unsafe development, deployment, or use of AI systems may pose catastrophic or even existential risks to humanity within our lifetimes}. These risks from misuse and loss of control could increase greatly as digital intelligence approaches or even surpasses human intelligence.” (International Dialogues on AI Safety, 2024)\footnote{Major signatories to the IDAIS statement include: Andrew Yao (Dean of Institute for Interdisciplinary Information Sciences, Tsinghua University), Ya-Qin Zhang (Chair Professor of AI Science at Tsinghua University), HongJiang Zhang (Chairman of the Beijing Academy of AI), Stuart Russell (Electrical Engineering and Computer Science, UC Berkeley), and Yoshua Bengio (Professor, Department of CS and Operations Research, Université de Montréal).}
    
\item 
    “\textbf{Coordinated global action} on AI safety research and governance is critical to prevent uncontrolled frontier AI development from posing unacceptable risks to humanity.” (International Dialogues on AI Safety, 2023)

\item 
    “We call on leading AI developers to make a \textbf{minimum spending commitment} of one-third of their AI R\&D on AI safety and for government agencies to fund academic and non-profit AI safety and governance research in at least the same proportion.” (International Dialogues on AI Safety, 2023).

\item 
    “\textbf{Safe, secure and trustworthy artificial intelligence systems}... are such that they are human-centric, reliable, explainable, ethical, inclusive, in full respect, promotion and protection of human rights and international law, privacy-preserving, sustainable development-oriented, and responsible.” (United Nations General Assembly, 2024).
    
\end{itemize}

\subsubsection{International cooperation and global governance}

Joint statements including representatives from the United States and China have also recognized the importance of global cooperation to avoid some of the most catastrophic outcomes of advanced AI development. Such statements have emphasized that risks arising from AI are “inherently international in nature” and suggested that governments should work together to define “clear red lines that, if crossed, mandate immediate termination of an AI system.”

Example quotes:
\begin{itemize}
\item
    “In the depths of the Cold War, international scientific and governmental coordination helped avert thermonuclear catastrophe. \textbf{Humanity again needs to coordinate to avert a catastrophe that could arise from unprecedented technology}.” (International Dialogues on AI Safety, 2024).
    
\item
    “\textbf{Many risks arising from AI are inherently international in nature}, and so are best addressed through international cooperation.” (Bletchley Declaration, 2023).
    
\item
    “We also recommend defining \textbf{clear red lines that, if crossed, mandate immediate termination of an AI system — including all copies — through rapid and safe shut-down procedures}. Governments should cooperate to instantiate and preserve this capacity. Moreover, prior to deployment as well as during training for the most advanced models, developers should demonstrate to regulators' satisfaction that their system(s) will not cross these red lines.” (International Dialogues on AI Safety, 2023)
\end{itemize}

\section{Discussion}
We compiled statements about advanced AI discourse in China and the United States, with a focus on statements about safety and security risks and international cooperation. Overall, major technical officials and policymakers in both countries have acknowledged extreme risks from advanced AI development. More specifically, experts in both China and the United States seem to have a shared understanding of some of the concrete safety and security challenges. For example, top Chinese scientists discuss concepts like “intelligence explosions”, risks from systems that can “self-replicate or self-modify”, and risks from systems that “do not hesitate to harm human beings” in order to survive. American scientists and policymakers have also acknowledged these risks, noting that AI could produce “loss of control” scenarios and calling for more work on “the risks that AGI development would pose.” 

There were also some noteworthy differences. For example, the Chinese discourse includes discussions about how AI may produce content that does not conform to the Chinese Communist Party's restrictions on speech. Experts have discussed how AI might help users generate “content that violates Socialist values”, such as content that can “incite the subversion of state power” or “overthrow the Socialist System.” Chinese standards for AI security also note that users should be suspended if they try to generate “illegal, unhealthy information.” To highlight another difference, the United States discourse included details about the role of the US AI Safety Institute in advancing the science of AI safety and setting international standards. Several countries– including the United States and United Kingdom– recently agreed to establish an international network of AI safety institutes to “forge a common understanding of AI safety” and share information about model capabilities and risks (UK Government, 2024). So far, China is not part of this agreement and has not yet announced plans for a formal AI safety institute.

Our work should be interpreted in light of a few important limitations. First, our work did not aim to be a comprehensive summary of AI discourse. We focused on a particular subset of the discourse (discourse related to safety and security concerns around advanced AI development) among a particular group of stakeholders (policymakers and technical experts who inform policymakers). Future work could examine discourse around other topics (e.g., bias and fairness, US-China competition, military applications of AI) among a wider array of stakeholders (e.g., members of the general public, investors, civil society groups). Second, our methodology was subjective– we did not aim to perform a comprehensive literature review or statistically examine trends in the entire corpus of discourse. Rather, we aimed to provide brief and illustrative findings that can offer readers a general understanding of the discourse. Future work could aim to be more comprehensive or quantitative. Third, our work only included publicly available statements– we do not provide insights into private discussions among policymakers. Finally, it should be noted that AI discourse can change rapidly in response to new events. As AI progress continues and policymakers' understanding of AI increases, we should suspect important changes in how policymakers in each country view the safety and security risks. 

Overall, we hope our work provides some useful insights into the state of advanced AI discourse in China and the United States. Such work could contribute to a broader body of literature that helps policymakers better prepare for the global security implications of advanced AI and inform discussions about global governance strategies. 

\pagebreak
\end{CJK}

\bibliographystyle{apalike}
\bibliography{example}
\sloppy

Amodei, Dario. (2023). Written Testimony for a hearing on Oversight of A.I.: Principles for Regulation. \url {https://www.judiciary.senate.gov/imo/media/doc/2023-07-26_-_testimony_-_amodei.pdf} 

Bengio, Yoshua. (2023). Written Testimony for a hearing on Oversight of A.I.: Principles for Regulation. \url {https://www.judiciary.senate.gov/imo/media/doc/2023-07-26_-_testimony_-_bengio.pdf} 

Bipartisan Senate AI Working Group. (2024). A Roadmap for Artificial Intelligence Policy in the United States Senate. \url {https://www.schumer.senate.gov/imo/media/doc/Roadmap_Electronic1.32pm.pdf} 

Bletchley Declaration. (2023). The Bletchley Declaration by Countries Attending the AI Safety Summit, 1-2 November 2023. \url{https://www.gov.uk/government/publications/ai-safety-summit-2023-the-bletchley-declaration/the-bletchley-declaration-by-countries-attending-the-ai-safety-summit-1-2-november-2023}

Department for Science, Innovation and Technology. (2024). Global leaders agree to launch first international network of AI Safety Institutes to boost cooperation of AI [Press release]. \url {https://www.gov.uk/government/news/global-leaders-agree-to-launch-first-international-network-of-ai-safety-institutes-to-boost-understanding-of-ai}

Embassy of the People’s Republic of China. (2023). Global AI Governance Initiative. \url{http://gd.china-embassy.gov.cn/eng/zxhd_1/202310/t20231024_11167412.htm}

Fang, Bingxing. (2020). Artificial Intelligence Safety and Security. \url{https://chineseperspectives.ai/Binxing-FANG}

Guinness, Harry (2024). The Best Large Language Models of 2024. \url{https://zapier.com/blog/best-llm/}

International Dialogues on AI Safety. (2023). IADS Ditchley Park 2023. \url{https://idais.ai/}

Henshall, W. (2024, February 21). How a New Bipartisan Task Force is Thinking About Artificial Intelligence. TIME. \url {https://time.com/6727264/house-artificial-intelligence-task-force/} 

International Dialogues on AI Safety. (2024). IADS Beijing 2024. \url{https://idais.ai/}

Ministry of Foreign Affairs of the People’s Republic of China. (2024). Wang Yi on Global AI Governance: Ensure that AI is a Force for Good, Ensure Safety and Ensure Fairness. \url{https://www.fmprc.gov.cn/eng/zxxx_662805/202403/t20240308_11256430.html#:~:text=Wang%20Yi%20said%20artificial%20intelligence,be%20checked%20before%20setting%20off}

Statista (2024). Artificial Intelligence - United States. \url{https://www.statista.com/outlook/tmo/artificial-intelligence/united-states}

National Institute of Standards and Technology. (2024). The United States Artificial Intelligence Safety Institute: Vision, Mission, and Strategic Goals. \url {https://www.nist.gov/system/files/documents/2024/05/21/AISI-vision-21May2024.pdf} 

Romney et. al. (2024). Framework for Mitigating Extreme AI Risks. \url {https://www.romney.senate.gov/wp-content/uploads/2024/04/AI-Framework_2pager.pdf} 

Senate Committee on Commerce, Science, and Transportation. (2024). The Future of AI Innovation Act of 2024: Section-by-Section Summary. \url {https://www.commerce.senate.gov/services/files/DE2C57C7-7F3C-4D0F-9DA3-FB66E15E9A8A} 

Statista (2024). Artificial Intelligence - China. \url{https://www.statista.com/statistics/1262377/china-ai-market-size/}
TC260. (2024). Basic Security Requirements for Generative Artificial Intelligence Services. \url{https://www.tc260.org.cn/upload/2024-03-01/1709282398070082466.pdf}

United Nations General Assembly. (2024). A/78/L.49. \url{https://documents.un.org/doc/undoc/ltd/n24/065/92/pdf/n2406592.pdf?token=PWMWxny2akiFYYDipm&fe=true}

Yang, Zeyi. (2024). Four things you need to know about China’s AI talent pool. MIT Technology Review. \url{https://www.technologyreview.com/2024/03/27/1090182/ai-talent-global-china-us/#:~:text=1.,industry%20to%20absorb%20that%20talent.%E2%80%9D}

Yuqing, et. al. (2021),  Technical Countermeasures for Security Risks of Artificial Intelligence. Strategic Study of CAE 2021, Volume 23, Issue 3. \url{https://www.engineering.org.cn/en/article/29491/detail}

Zeng, Yi. (2023). Chinese Perspectives on AI Safety. \url{https://chineseperspectives.ai/yi-zeng}

Zhang, Bo. (2022). Academician Zhang Bo: Make Responsible AI. \url{https://chineseperspectives.ai/Bo-ZHANG}

Zhang, Irene. (2023). Putting China’s Top LLMs to the Test. Chinatalk. \url{https://www.chinatalk.media/p/putting-chinas-top-llms-to-the-test}

Zhang, Ya-Qin. (2023). Embracing AI by Prioritizing Values over Technology. Internet Weekly, Issue 15, 2023. \url{https://chineseperspectives.ai/Ya-qin-Zhang}
\end{document}